\documentclass[aps,prl,twocolumn,showpacs]{revtex4}
\usepackage{amssymb}
\usepackage{amsmath}
\usepackage{epsfig}

\begin{document}
\title{
Landau-Zener Bloch oscillations with perturbed flat bands}

\author{Ramaz Khomeriki$^{1,2}$, Sergej Flach$^{2,3}$}
\affiliation {
${\ }^1$Physics Department, Tbilisi State University, Chavchavadze
3, 0128 Tbilisi, Georgia \\
${\ }^2$Center for Theoretical Physics of Complex Systems, Institute for Basic Science, Daejeon, South Korea \\
${\ }^3$New Zealand Institute for Advanced Study,
Centre for Theoretical Chemistry \& Physics, Massey University,
Auckland, New Zealand
}

\begin{abstract}
Sinusoidal Bloch oscillations appear in band structures exposed to
external fields. Landau-Zener (LZ) tunneling between different bands is
usually a counteracting effect limiting Bloch oscillations. Here we
consider a flat band network with two dispersive and one flat bands,
e.g. for ultracold atoms and optical waveguide networks. Using
external synthetic gauge and gravitational fields we obtain a
perturbed yet gapless band structure with almost flat parts. The
resulting Bloch oscillations consist of two parts - a fast scan
through the nonflat part of the dispersion structure, and an almost
complete halt for substantial time when the atomic wave packet is
trapped in the original flat band part of the unperturbed spectrum, made possible due to LZ tunneling.

\end{abstract}
\pacs{67.85.-d, 37.10.Jk, 03.65.Ge, 03.65.Aa} \maketitle

Waves  probe the symmetries and topologies imprinted by an
underlying periodic potential.Quite often it is possible to restrict
the dynamics to a few bands, for instance for electrons in crystals
or in artificial quantum dot arrays~\cite{mesoscopic_review},
ultracold atoms in optical lattices~\cite{bloch2008}, microwaves in
dielectric resonator networks~\cite{bellec13}, and light propagation
in waveguide networks~\cite{christodoulides2003}.  Additional
interactions between the constituent waves lead to interesting new
phenomena (see references in recent reviews on topological flat
bands~\cite{bergholtz13,parameswaran2013}).

Flat band (FB) networks are specific tight-binding translationally
invariant lattices with local symmetries which ensure the existence
of one (or a few) completely dispersionless bands in the spectrum.
FBs have been studied in a number of lattice models in
three-dimensional, two-dimensional, and even one-dimensional (1D)
settings~\cite{richter,hyrkas13,derzhko_review}, and recently
realized experimentally with photonic waveguide
arrays~\cite{moti_lieb,compact_lieb_1,compact_lieb_2},
exciton-polariton
condensates~\cite{masumoto2012,flat_exciton_polariton}, and
ultra-cold atomic condensates~\cite{optical_lieb}.

FB networks rely on the existence of compact localized eigenstates (CLS) due to destructive interference,
enabled by the local symmetries of the network. FB networks are constructed using graph theory
~\cite{mielke,tasaki,richter,dias2015,nandy2015}, CLS ~\cite{bergman2008,richter,flach14}, and can be perturbed e.g. by
disorder to arrive at
unexected new scaling laws \cite{flach14,leykam16}, and correlated potentials to arrive at
diverging densitites of states, gaps, and designable mobility edges \cite{bodyfelt14,danieli15}.
A perhaps most prominent example is the
fractional quantum Hall effect, which occurs as a result of the flat
band degeneracy of Landau levels of electrons in a magnetic field and e-e interactions
\cite{storm}.

A very intriguing question which might be important for applications
is whether and how compact localized states will perform Bloch
oscillations in the presence of external fields. We choose a FB
network which has an easily realizable geometry, and a flat band
which intersects dispersive bands. Following \cite{flach14} we
decide to perturb the FB of the diamond chain network \cite{vidal00}
and hybridize it with dispersive modes under the action of dc
electric and magnetic fields, although we could as well use the
one-dimensional Lieb lattice \cite{flach14} for our purposes. Our
results are applicable for ultracold atoms in optical lattices where
the electric field is substituted by a tilt of the lattice in the
gravitational field \cite{kasevich} or accelerating the whole
lattice \cite{camp}, while the magnetic field is generated by
artificial gauge fields \cite{gauge}. Notably the same type of
perturbations can be arranged  in optical waveguide arrays where the
electric field is modeled by a curved geometry of the waveguides
\cite{longhi}, while a special metallic fabrication of the
waveguides and the surrounding medium \cite{metal} mimics a magnetic
flux.
\begin{figure}[t]
\includegraphics[scale=.5]{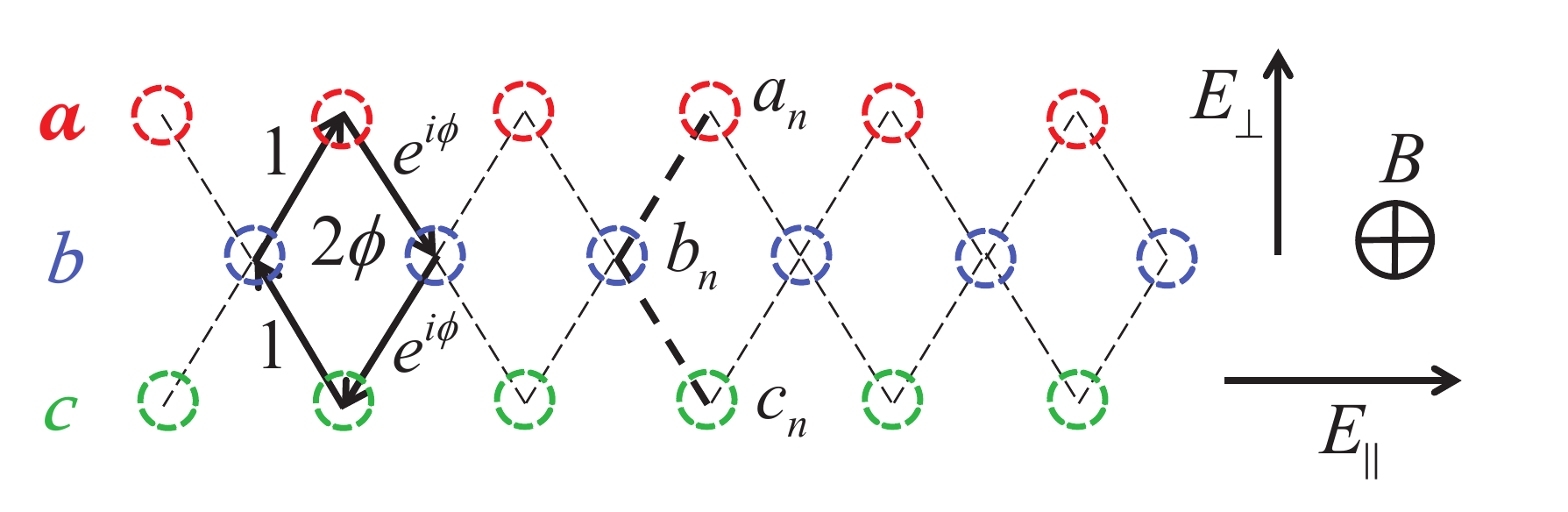}
\caption{a) Schematics for the three leg diamond lattice with $a$,
$b$ and $c$ legs. Dashed lines indicate sites connected with hopping
(tunneling) of the particle/wave. Solid arrows indicate the phase of
complex hopping constants in a single plaquette, with the specific
gauge used for the perpendicular (to the lattice plane) dc magnetic
field $B$. $E_\parallel$ and $E_\perp$ define the longitudinal and
transversal components of the dc electric field, respectively.}
\label{fig1}
\end{figure}
The applied dc field only partly removes the flatness of the FB
leading to a gapped band structure. A properly added magnetic flux
closes these gaps and enforces Landau-Zener tunneling \cite{zener}
which scans the whole band structure in Bloch oscillation manner
\cite{osc}, and comes to an almost complete halt once the wave is
exploring the reminders of the unperturbed flat band.

\begin{figure}[t]
\includegraphics[width=\columnwidth]{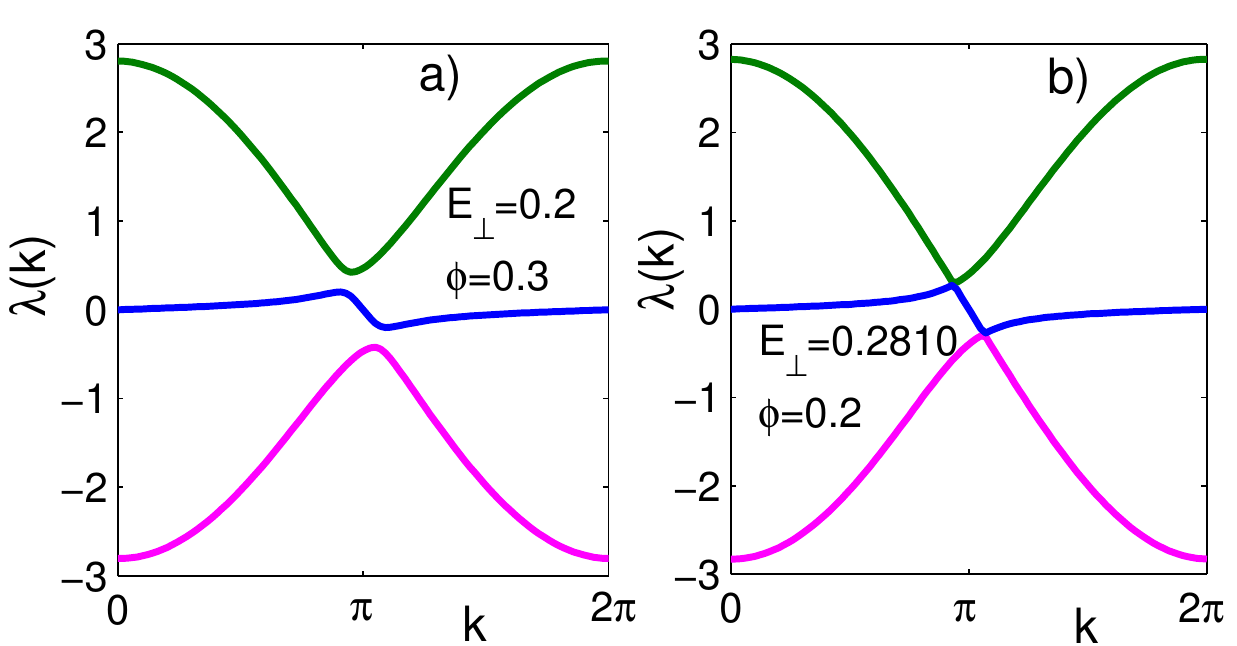}
\caption{Band structure for $E_{\parallel}=0$ and different values
of transversal fields: a) represents the typical situation of
avoided crossings for $\phi=0.3$ and $E_{\perp}=0.2$. b) displays a
complete gap closing case for $\phi=0.2$ and
$E_{\perp}=\sqrt{2}\sin\phi$.} \label{fig2}
\end{figure}

The tight-binding Hamiltonian of the diamond chain reads as follows:
\begin{eqnarray}
&{\cal \hat H}={\cal \hat H}_{h}+{\cal \hat
H}_\perp+{\cal \hat H}_\parallel, \label{1} \\
&{\cal \hat H}_{h}= -\sum(\hat b_n^+\hat a_n+\hat
c_{n}^+\hat b_{n})e^{i\phi}-\hat b_{n-1}^+\hat
c_n-\hat a_n^+\hat b_{n-1}+ c.c. \;,\nonumber \\
&{\cal \hat H}_\perp=\sum E_\perp\left(\hat
a_{n}^+\hat a_n-c_{n}^+\hat c_n\right), \nonumber
\\
&{\cal \hat
H}_\parallel=\sum E_{\parallel}n\left(\hat
a_{n}^+\hat a_n+\hat b_{n}^+\hat b_n+\hat c_{n}^+\hat
c_n\right)+\frac{E_\parallel}{2}\hat b_{n}^+\hat b_n \;,\nonumber
\end{eqnarray}
where $\hat a_{j}^+$, $\hat b_{j}^+$, $\hat c_{j}^+$ and $\hat
a_{j}$, $\hat b_{j}$, $\hat c_{j}$ are standard creation and
annihilation operators of an atom at the  $j$-th lattice site of the
legs $a$, $b$, $c$, but could also simply be complex amplitudes of a
photonic light field in a waveguide structure. $E_\parallel$ and
$E_\perp$ are longitudinal and transversal in-plane components of a
dc electric (gravitational) field and the phase $\phi$ complexifies
the tunneling amplitudes as a particular gauge choice for the dc
magnetic (artificial gauge) field $B$ which is oriented
perpendicular to the diamond chain embedding plane. It follows that
the magnetic flux penetrating each diamond plaquette has twice the
value - $2\phi$.

The discrete Schr\"odinger equation which follows from
\eqref{1} is given by
\begin{eqnarray}
&i\dot{a}_n=\left(E_\parallel n+E_\perp\right)
a_n-e^{-i\phi}b_n-b_{n-1} \;,
\nonumber \\
&i\dot{b}_n=E_\parallel\left(n+\frac{1}{2}\right)b_n-e^{i\phi}a_n-e^{-i\phi}c_n-c_{n+1}-a_{n+1} \;,
\nonumber \\
&i\dot{c}_n=\left(E_\parallel n
-E_\perp\right)c_n-e^{i\phi}b_n-b_{n-1}
\;.
\label{2}
\end{eqnarray}

In the absence of a longitudinal field $E_\parallel = 0$ we seek for plane wave solutions
$a_n=a(t)e^{ikn}$, $b_n=b(t)e^{ikn}$,
$c_n=c(t)e^{ikn}$:
\begin{eqnarray}
&i\dot{a} = E_\perp a-\left(e^{-i\phi}+e^{-ik}\right)b\;,
\nonumber \\
&i\dot{b} = -\left(e^{i\phi}+e^{ik}\right)a-\left(e^{-i\phi}+e^{ik}\right)c \;,
\nonumber \\
&i\dot{c}=
-E_\perp c-\left(e^{i\phi}+e^{-ik}\right)b \;, \label{3}
\end{eqnarray}
and arrive at the following cubic equation for the eigenvalue $\lambda$ using $a(t),b(t),c(t) \sim e^{i\lambda t}$:
\begin{eqnarray}
&-\lambda^3 + C\lambda - D = 0\;, \label{301}
\\
&C=E_{\perp}^2+4(1+\cos \phi \cos k), \quad D=4E_{\perp} \sin \phi
\sin k. \nonumber
\end{eqnarray}

For special cases it follows that one central (at $\lambda=0$) or
all three bands are flat \cite{vidal00}:
\begin{eqnarray}
\phi=0 \; : \; \lambda_1 = 0, \quad \lambda_\pm=\pm\sqrt{E_{\perp}^2+4(1+\cos k)} \quad
 \;, \nonumber
\\
E_{\perp}=0 \;,\; \phi=\pi/2 \; : \;
\lambda_1=0\;,\; \lambda_\pm = \pm 2\;,
\label{specialcases}
\\
E_{\perp}=0 \; : \;
\lambda_1=0 \;,\; \lambda_\pm = \pm 2 \sqrt{1+\cos \phi \cos k}
\;. \nonumber
\end{eqnarray}
Apart from those cases, all bands are nonflat, with a typical
dispersion shown in Fig. \ref{fig2}a. The central
flatband becomes dispersive, and we observe two gaps (avoided crossings)
symmetrically located around $k=\pi$ and $\lambda=0$.
\begin{figure}[b]
\includegraphics[width=\columnwidth]{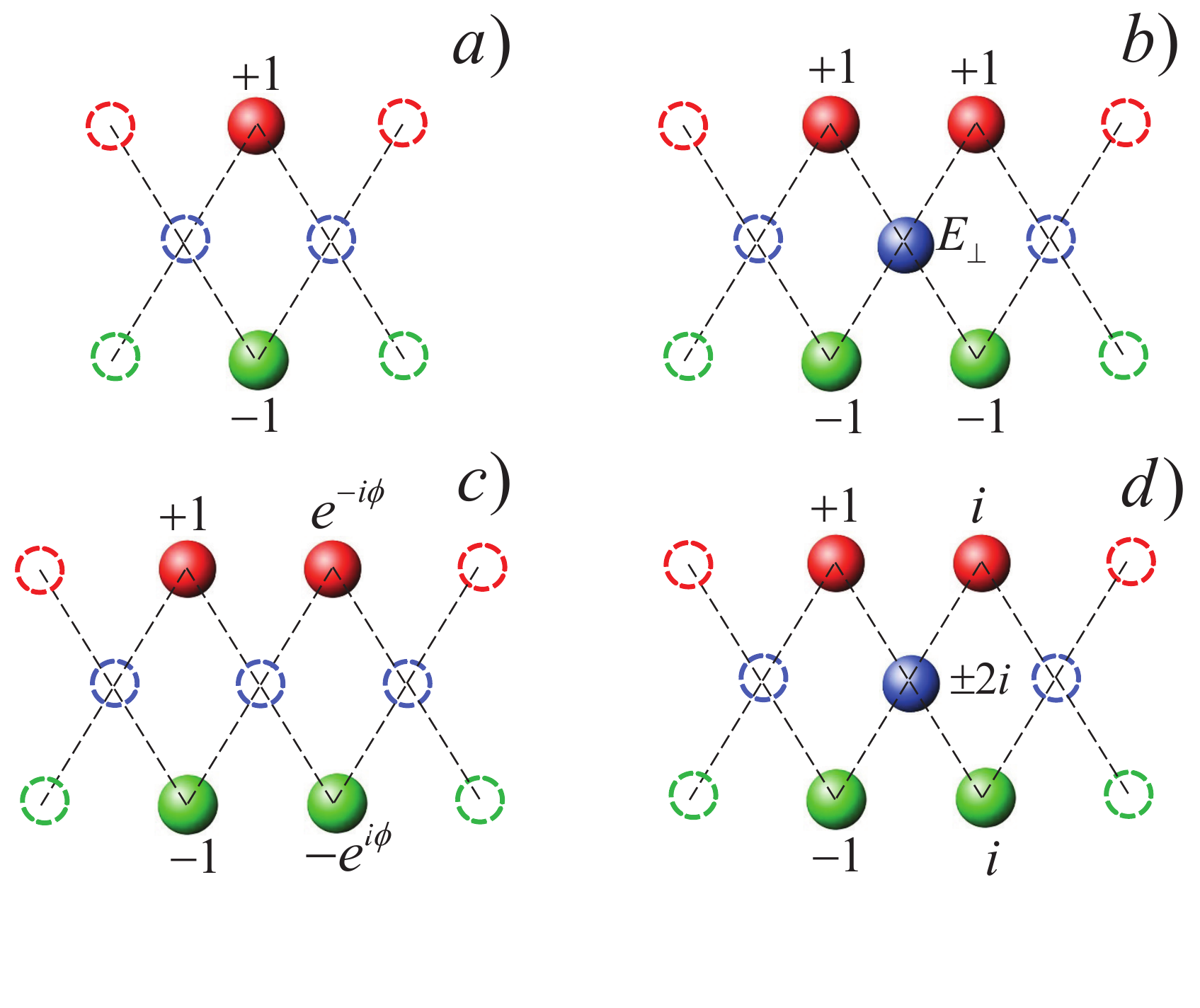}
\caption{CLS structure. Filled circles - excited sites with
amplitudes denoted right to them. Empty circles correspond to zero
amplitude. Lines indicate the hopping connections. (a)
$E_\perp=\phi=0$, $\lambda=0$; (b) $\phi=0$, $\lambda=0$; (c)
$E_\perp=0$, $\lambda=0$; (d) $E_\perp=0$, $\phi=\pi/2$, $\lambda =
\pm 2$. } \label{fig3}
\end{figure}

The Bloch eigenstates of a flat band can be superposed in order to
obtain compact localized eigenstates \cite{flach14}. Different
flatband networks are characterized by different local symmetries
and topologies of CLS. In one-dimensional settings the CLS can be
classified by the integer number $U$ of unit cells they occupy
\cite{flach14}. For $U=1$ CLS form an orthogonal linearly
independent set, while for $U > 1$ their set is linearly independent
but non-orthogonal. For a given network the value of $U$ can change
upon lowering local symmetries, e.g. due to external fields. For
$E_\perp=\phi=0$ the diamond chain belongs to the $U=1$ class, and
its compact localized state is shown in Fig.\ref{fig3}(a). Only two
sites within one unit cell are excited. Destructive interference
prevents a leaking of the wave functions into the exterior. For all
other flat band cases considered in \eqref{specialcases} the CLS
class increases to $U=2$. For $\phi=0$ the CLS at $\lambda=0$ is
shown in Fig.\ref{fig3}(b). Note that this CLS will tend to a single
site $U=1$ CLS in the limit of infinite $E_\perp$. For $E_\perp=0$
and nonzero flux $\phi$ the CLS vector becomes complex as shown in
Fig.\ref{fig3}(c) at $\lambda=0$. For the special case $\phi=\pi/2$
all three bands turn flat \cite{vidal00} and the new flat band
energies $\lambda = \pm 2$ correspond to the CLS vectors shown in
Fig.\ref{fig3}(d).
\begin{figure}[t]
\includegraphics[scale=.7]{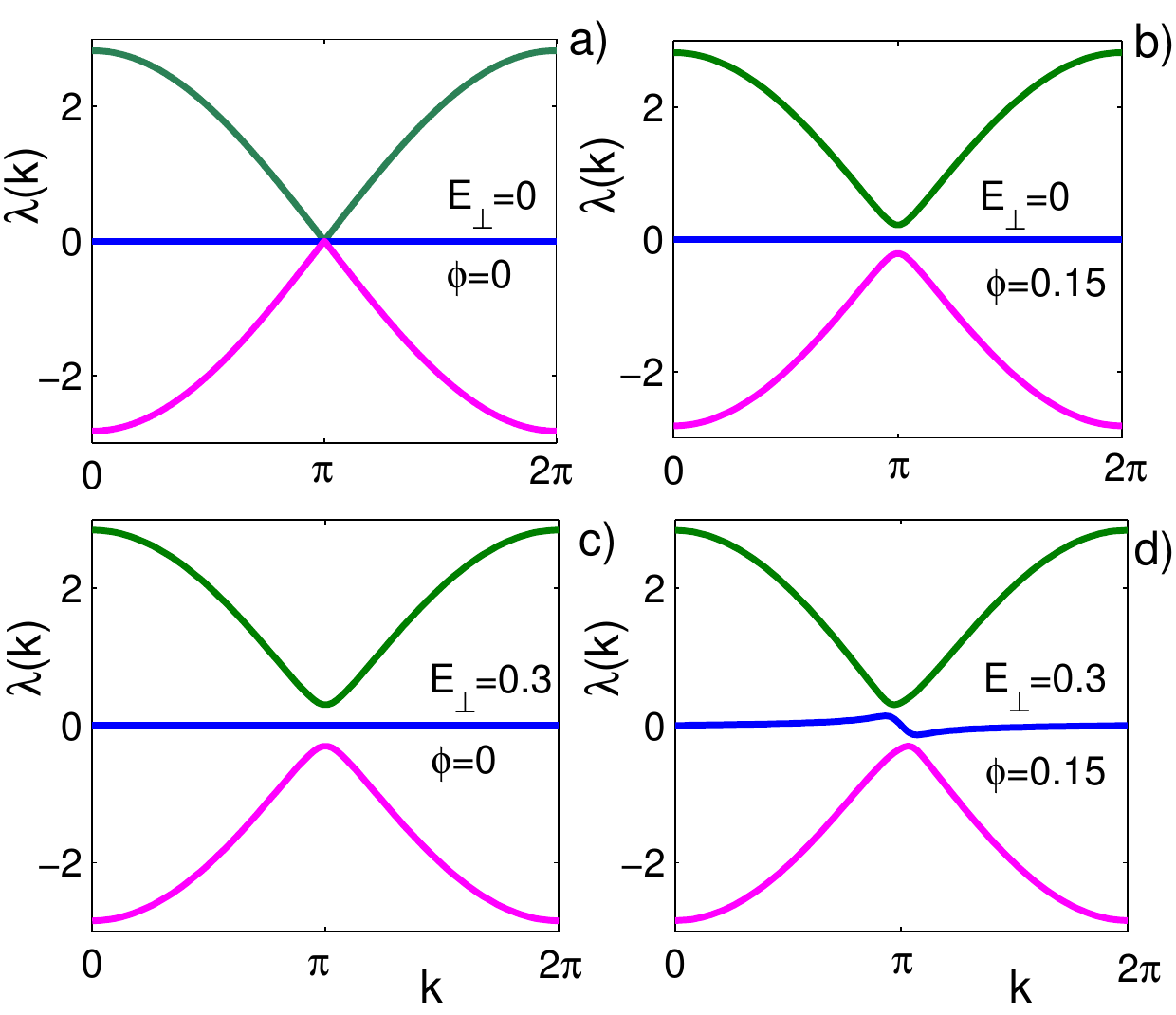}
\caption{Band energies versus wave number. Graph a) displays the case
when transversal fields are absent, in b) and c) one of the fields
is present, while in graph d) both dc electric and magnetic fields
are nonzero. The parameter values are indicated in the graphs. }
\label{fig4}
\end{figure}

By finetuning the parameters $E_\perp$ and $\phi$ we can close the
gaps in the band structure completely. Indeed, the cubic equation
(\ref{301}) has two degenerate roots when the Cardano equality $4C^3
= 27D^2$ is satisfied, which translates into
\begin{equation}
E_{\perp}^2+4(1+\cos \phi \cos k)=(6/2^{1/3})\left(E_{\perp} \sin
\phi \sin k\right)^{2/3} \;. \label{303}
\end{equation}
Let us take $\phi$ as the free variable. It is straightforward to
show that there is a solution to \eqref{303} wich is given by
\begin{equation}
E_{\perp}=\sqrt{2}\sin \phi\;,\; k=\pi\pm\phi \;.
\end{equation}
The band structure with vanishing gaps
for the particular value of $\phi=0.2$ is shown in Fig.
\ref{fig2}b. The perturbed flat band part still displays a significant portion which is almost dispersionless.
\begin{figure}[t]
\includegraphics[scale=.66]{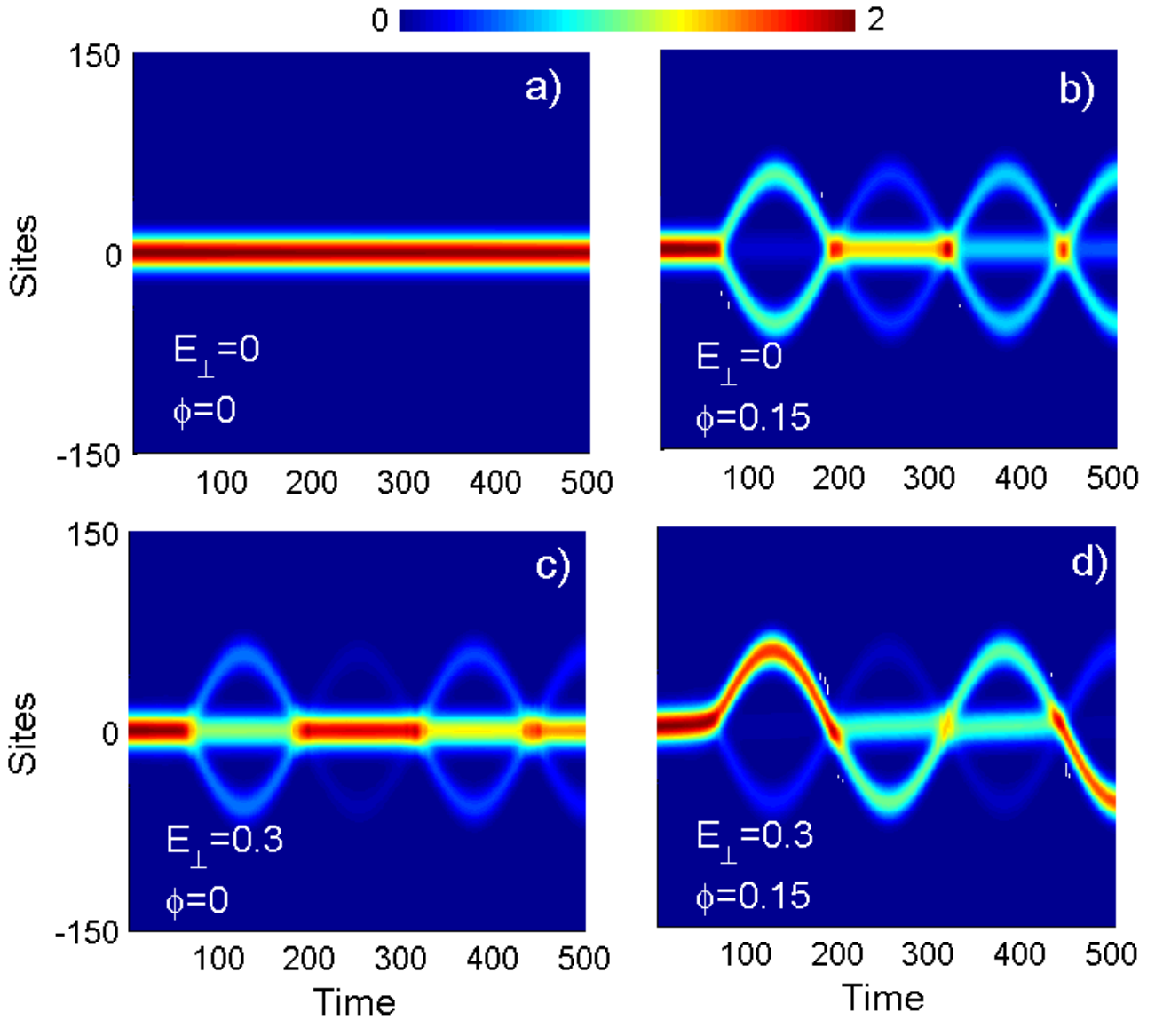}
\caption{Results of numerical simulations exciting initially the CLS
in Fig.\ref{fig3}(a), with respective field values as taken in graphs
of Fig. \ref{fig4}. In all simulations $E_\parallel =0.05$. The space-time evolution of the
norm density $\rho_n=|a_n|^2+|b_n|^2+|c_n|^2$ is plotted in color
code. } \label{fig5}
\end{figure}

Next we consider the case $E_\parallel\neq 0$. We note that the CLS
for $E_\perp=\phi=0$ in Fig.\ref{fig3}(a) is still an exact solution
of the wave equations in the presence of nonzero $E_\parallel$, and
any linear combination of multiple CLS as well \cite{supp}. We will
therefore first consider a Gaussian wavepacket of such linear
combinations this state $a_n=-c_n={\rm e}^{-n^2/2\sigma^2}$ as an
initial state with variance $\sigma=70$, and trace its evolution for
$E_\parallel = 0.05$. Fig.\ref{fig4} shows the band structure for
four different parameter cases, and Fig.\ref{fig5} the evolution of
the norm density per unit cell $\rho_n=|a_n|^2+|b_n|^2+|c_n|^2$. In
all considered cases we observe Bloch oscillations. However, the
details of these oscillations are strongly depending on the
different cases considered.

The above mentioned case $E_\perp=\phi=0$ with a flat band at
$\lambda = 0$ (Fig.\ref{fig4} (a)) keeps the compactness of the
initial state, therefore no oscillations occur (Fig.\ref{fig5} (a)).
For $E_\perp = 0$ and $\phi \neq 0$ and similarly for $E_\perp \neq
0$ and $\phi=0$ the initial state starts to evolve in a symmetric
way (Fig.\ref{fig5}) (b,c) reflecting the band structure symmetry in
Fig.\ref{fig4} (b,c). We observe sharp changes from an almost frozen
state into an oscillating pattern due to Landau-Zener transitions at
the gaps of the band structure. These sudden switches from a
nonmoving to a rapidly oscillating packet are further intensified
and become asymmetric when both fields $E_\perp$ and $\phi$ turn
nonzero, as seen in Fig.\ref{fig5} (d) respectively. Still the
nonzero gap values induce a splitting of the wave packet - a part of
the packet continues adiabatically, while a complementary part
continues anti-adiabatically due to Landau-Zener transitions.

Let us quantify these observations. Since we choose longitudinal
field values $E_\parallel = 0.05$ which are small compared to the
band structure width $2\sqrt{E^2_\perp +8}$ we can describe the
Bloch oscillations by replacing $k$ in the band structure for
$E_\parallel = 0$ with the slow variable $E_\parallel t$ which scans
the band structure. The starting point corresponds to the (almost)
flat band in Fig.\ref{fig4}(b-d). The avoided crossings are reached
at $k=\pi$ which translated into $t=\pi/E_\parallel = 63$ in
excellent agreement with Fig.\ref{fig5}(b-d). The gap values are
$\Delta=0.18\;,\;0.32\;,\;0.16$ for the cases (b,c,d) in
Fig.\ref{fig4} respectively. The  relevant Landau-Zener parameter
$\alpha$ which describes the scanning speed through the avoided
crossing is given by $\alpha=E_\parallel \sqrt{2} \cos ^2(\phi/2)$
\cite{supp}. The probability $P = {\rm e}^{-\pi \Delta^2 / 2\alpha}$
of a diabatic Landau-Zener transition \cite{zener} is then obtained
as $P=0.48 \;,\; 0.1 \;,\; 0.56$ for the cases (b,c,d) in
Fig.\ref{fig5}. Indeed, the numerically observed diabatic
transitions are much stronger in cases (b,d) and weaker in case (c)
in Fig.\ref{fig5}.

\begin{figure}[t]
\includegraphics[scale=.7]{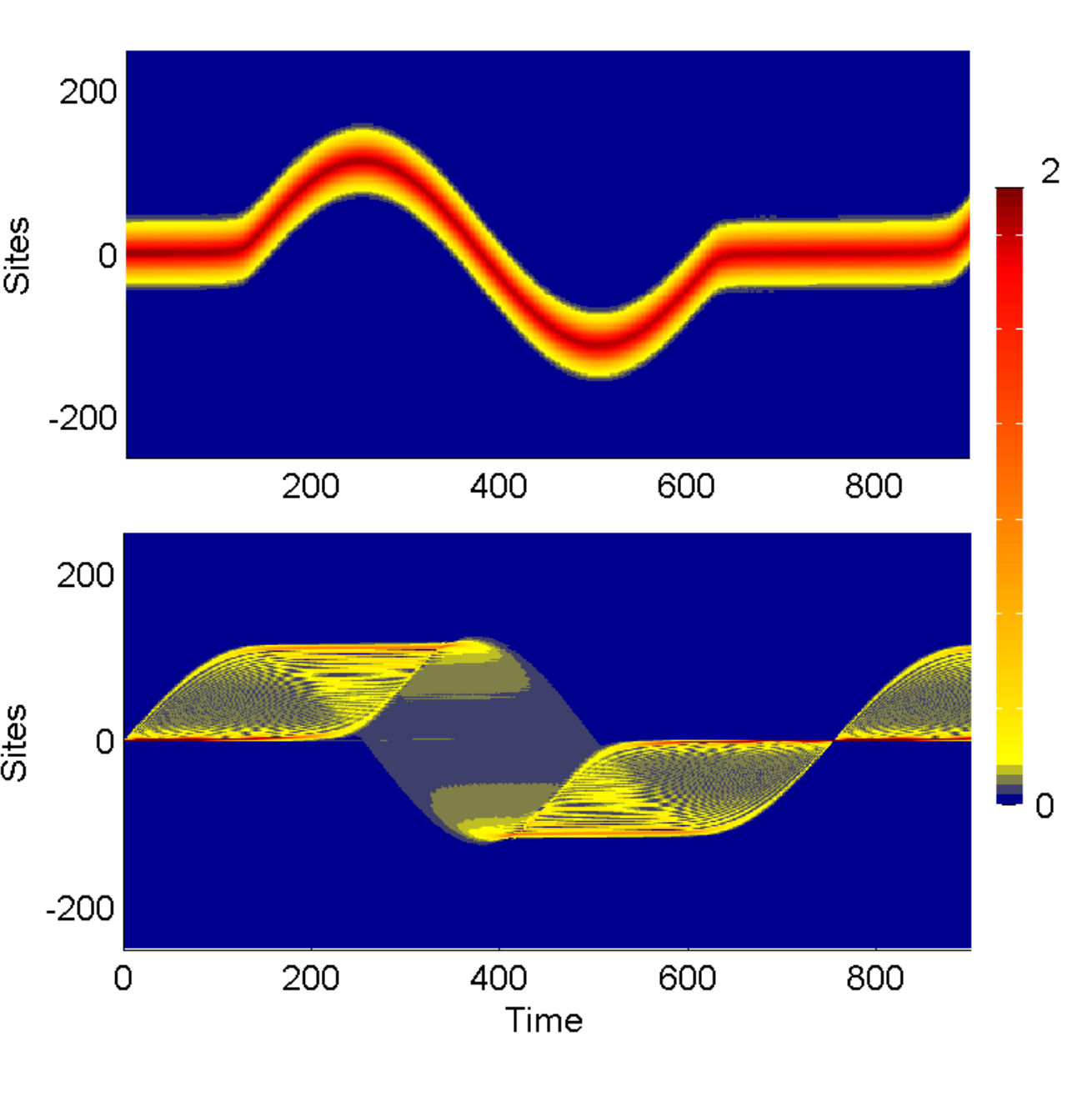}
\caption{The case of a gapless band structure is simulated for
$\phi=0.1$, $E_\perp=\sqrt{2}\sin\phi$ and $E_\parallel=0.025$.
The
plots display the space-time evolution of the norm density
$\rho_n=|a_n|^2+|b_n|^2+|c_n|^2$.
Upper plot: initial state is Gaussian as before. Lower plot: Initial state is in the form of
a single CLS
as in Fig.\ref{fig3}(a).
 } \label{fig6}
\end{figure}

We proceed to the case of vanishing gaps as discussed above.
The most striking impact of both transversal dc
electric and magnetic fields is that for
$E_\perp=\sqrt{2}\sin\phi$, when the band structure turns gapless,
Bloch oscillations harvest completely from Landau-Zener tunneling whose probability turns into unity.
This case is displayed in the upper plot in Fig. \ref{fig6}.
For a significant part of
its evolution it stands still, only to cross over into a large amplitude oscillation which clearly corresponds to the
scanning of the corresponding band structure with complete Landau-Zener tunneling.
These features are observed even in the case of an initial condition in the form a single CLS from Fig.\ref{fig3}(a) and shown in the lower plot in Fig.\ref{fig6}.
In this case, the CLS remains essentially frozen until it performs a violent sweep through the system over about 30 sites, where it recombines only to perform the sweep again.

The combination of the physics of Bloch oscillations, Landau-Zener tunneling and of flat band networks opens promising directions for control of unconventional quantum transport
through properly designed lattice structures. Extensions of this study to two dimensions and to the inclusion of two-body interactions are intriguing pathways for further work.

\end{document}